\documentclass[doublecol]{epl2}
\usepackage{graphicx}% Include figure files
\usepackage{dcolumn}% Align table columns on decimal point
\usepackage{bm}% bold math
\usepackage{epsf}
\usepackage{subfigure}
\usepackage{amsmath}
\usepackage{amssymb}
\usepackage[cspex,bbgreekl]{mathbbol}
\usepackage{color}
\usepackage{stackrel}
\usepackage[T1]{fontenc}

\renewcommand{\i}{{\rm i}}

\newcommand{\be}{\begin{equation}}
\newcommand{\ee}{\end{equation}}
\newcommand{\ba}{\begin{eqnarray}}
\newcommand{\ea}{\end{eqnarray}}
\newcommand{\s}{\boldsymbol \sigma}
\newcommand{\M}{{\bf M}}
\newcommand{\m}{{\bf m}}
\newcommand{\Mb}{\overline{{\bf M}}}

\newcommand{\B}{{\bf B}}

\begin{document}

\title{Thermodynamic bounds on equilibrium fluctuations of a global or local order parameter}

\author{J. Guioth\inst{1} \and D. Lacoste\inst{1}}

\institute{                    
  \inst{1} Laboratoire de Physico-Chimie Th{\'e}orique - UMR CNRS Gulliver 7083,\\ PSL Research University, ESPCI, 
10 rue Vauquelin, F-75231 Paris, France
}

\date{\today}

\abstract{We analyze thermodynamic bounds on equilibrium fluctuations of an order parameter,
which are analogous to relations, which have been derived 
recently in the context of non-equilibrium fluctuations of currents.  
We discuss the case of {\it global} fluctuations when the order parameter is 
measured in the full system of interest, and {\it local} fluctuations, 
when the order parameter is evaluated only in  
a sub-part of the system. Using isometric fluctuation theorems, 
we derive thermodynamic bounds on the
fluctuations of the order parameter in both cases.
These bounds could be used to infer the value of symmetry breaking field or the 
relative size of the observed sub-system to the full system from {\it local} fluctuations.}

\pacs{05.70.Ln}{Nonequilibrium and irreversible thermodynamics}
\pacs{05.40.-a}{Fluctuation phenomena, random processes, noise, and Brownian motion}
\pacs{05.70.-a}{Thermodynamics}

\maketitle
%%%%%%%%%%%%%%%%%%%%%%%%%%%%%%%%%%%%%%%%%%%%%%%%%%%%%%%%%%%%%%%%%%%%%

Recently, a set of thermodynamic bounds have been obtained, which 
have a linear-response form and express a trade-off between the variance of current fluctuations and the rate
of entropy production \cite{Barato2015,Pietzonka2016}. These relations contribute to the field of statistical kinetics  
and could represent important trade-offs in the design of living 
systems. Following this work, these uncertainty bounds have been derived 
rigorously from large deviation theory 
\cite{Gingrich2016}. Specific bounds on current fluctuations 
have also been obtained separately for the symmetric exclusion process and for 
diffusive systems \cite{Akkermans2013}.

Dissipative systems break the time-reversal symmetry; but 
the formalism of large deviation theory is general and is also applicable to equilibrium fluctuations \cite{Derrida2007_vol,Touchette2009}. 
For equilibrium fluctuations, other forms of symmetry breaking not related to time are known.  
For instance, an ensemble of $N$ Ising spins in a magnetic field is a classic illustration of an equilibrium system with 
discrete symmetry breaking. In discussing this pedagogical example \cite{Goldenfeld1992_vol}, N. Goldenfeld 
derived a simple relation for the ratio of the probability to observe a magnetization 
$\M_N$, $P_{\bf B}(\M_N)$ with the probability to observe instead $-\M_N$:
\be
P_{\bf B}(\M_N) = P_{\bf B}(-\M_N) \ {\rm e}^{2 \beta \B\cdot \M_N}. 
\label{FT1}
\ee
The similarity of Eq.~(\ref{FT1}) with the Gallavotti-Cohen fluctuation theorem    
has been briefly noticed in \cite{Kurchan2009} and only extensively studied in \cite{Gaspard2012}.
Inspired by these works and by the discovery of fluctuation relations combining spatial and time-reversal symmetries called 
isometric fluctuation relations \cite{Hurtado2011,Hurtado2014}, 
one of us derived an extension of Eq.~(\ref{FT1}) for general symmetries described by group theory \cite{Lacoste2014}, which was then 
illustrated on a number of classic models of statistical physics \cite{Lacoste2015a}.

In this paper, we derive analogs of the thermodynamic   
uncertainty bounds for equilibrium systems with symmetry breaking. 
Using Eq.~(\ref{FT1}), we find under some restrictive conditions to be detailed below, 
the following inequality for the variance of $\M_N$: 
\be
\frac{{\rm Var}(M_N)}{\langle M_N \rangle} \le \frac{k_B T}{B},
\label{inequality}
\ee
where we denote the projection of $\M_N$ along $\B$ as $M_N$.
In terms of the magnetization density $\m=\M_N/N$, 
this relation is equivalent to ${\rm Var}(m)/\langle m \rangle \le k_B T/ BN$. 

We will also prove a different inequality, which instead holds more generally 
for any system of classical spins (Ising or Heisenberg) and arbitrary couplings 
and reads:
\be
\frac{\langle M_N^2 \rangle}{\langle M_N \rangle} \ge \frac{k_B T}{B}.
\label{lower-bound}
\ee
A remarkable feature of the inequalities of Eqs.~(\ref{inequality})-(\ref{lower-bound}) is that they hold beyond the linear response regime of small $B$, 
but become saturated when $B \to 0$ \cite{Goldenfeld1992_vol}.

The outline of this paper is as follows: after providing an equivalent form of Eq.~(\ref{inequality}), we consider 
some simple cases with one or two spins, then 
we prove Eq.~(\ref{lower-bound}) for an ensemble of classical Heisenberg spins. 
The rest of the paper investigates the validity of Eq.~(\ref{inequality}) 
for large systems either using a {\it global} or a {\it local} order parameter. 

For a finite number of spins, the magnetic susceptibility $\chi_N=d \langle M_N \rangle/dB$ satisfies
the fluctuation-response relation $\chi_N = \beta {\rm Var}(M_N)$ for any finite value of the magnetic field \cite{Chaikin1995_vol}. 
Thus, the inequality of Eq.~(\ref{inequality})
is equivalent to 
\be
\chi_N \leq \frac{\langle M_N \rangle}{B}.
\label{chi_ineq}
\ee 
It is reasonable that such a relation should hold independently of the temperature 
because it holds at least near $B=0$ (where the inequality is saturated) and near $B \to \infty$. 
Indeed, in the latter case, the susceptibility vanishes due to the saturation of the magnetization, 
and the right hand side of Eq.~(\ref{chi_ineq}) also vanishes because the average magnetization 
is bounded and $B \to \infty$. 
Whether the inequality should hold also in the intermediate range of values of $B$ is the real question.

In order to investigate this, let us consider a simple case namely that of 
of a single Ising spin $\sigma=\pm 1$ at temperature $T$
in a magnetic field $B>0$.  Using the canonical distribution: $p(\sigma)=e^{\beta B \sigma}/Z$, with $Z$ the partition
function, it is straightforward to show that $\langle \sigma^2 \rangle=1$ and $\langle \sigma \rangle = \tanh (\beta B)$.
Our inequalities Eq.~(\ref{lower-bound}) and Eq.~(\ref{inequality}) are indeed verified since :
\be
\frac{ \langle \sigma^2 \rangle }{\langle \sigma \rangle} = \coth (\beta B) \ge \frac{k_B T}{B},
\label{ineq1}
\ee
and
\be
\frac{ \langle \sigma^2 \rangle - \langle \sigma \rangle^2}{\langle \sigma \rangle} = \frac{1 - \tanh (\beta B)^2}{\tanh (\beta B)} \le \frac{k_B T}{B}.
\label{ineq2}
\ee

Now, let us consider two such Ising spins $\sigma_1$ and $\sigma_2$ 
interacting with a coupling constant $J$ again at temperature $T$ 
and in a magnetic field $B>0$. Naturally, we are interested in the fluctuations of the 
total magnetization, $M_2=\sigma_1 + \sigma_2$. A straightforward
calculation gives 
\be
\frac{ \langle M_2^2 \rangle}{\langle M_2 \rangle}= 2 \coth (2 \beta B) \ge \frac{k_B T}{B}, 
\label{ineq1-2spins}
\ee
which confirms Eq.~(\ref{lower-bound}) for $N=2$. Now, 
\be
\frac{ {\rm Var} (M_2)}{\langle M_2 \rangle}=\frac{2 \left( \exp (2 \beta J) + \cosh (2 \beta B) \right)}{\sinh (2 \beta B) ( \exp (2 \beta J) 
\cosh (2 \beta B)+1 )}.
\label{ineq2-2spins}
\ee
By maximizing the function on the right hand side of this equation over $J$ with $J \ge 0$, one finds that the maximum is reached for $J=0$. 
Therefore, ${\rm Var}(M_2) / \langle M_2 \rangle \le  2/ \sinh(2 \beta B) \le k_B T/B$, which confirms Eq.~(\ref{inequality}) for $N=2$
in that case. 
However, note that if we allow 
antiferromagnetic couplings ($J<0$), Eq.~(\ref{inequality}) can be violated in some range of values of $B$. 

\begin{figure}[h!]
%\begin{center}
%\includegraphics[scale=0.4]{fig-chi.pdf}
\onefigure[scale=0.4]{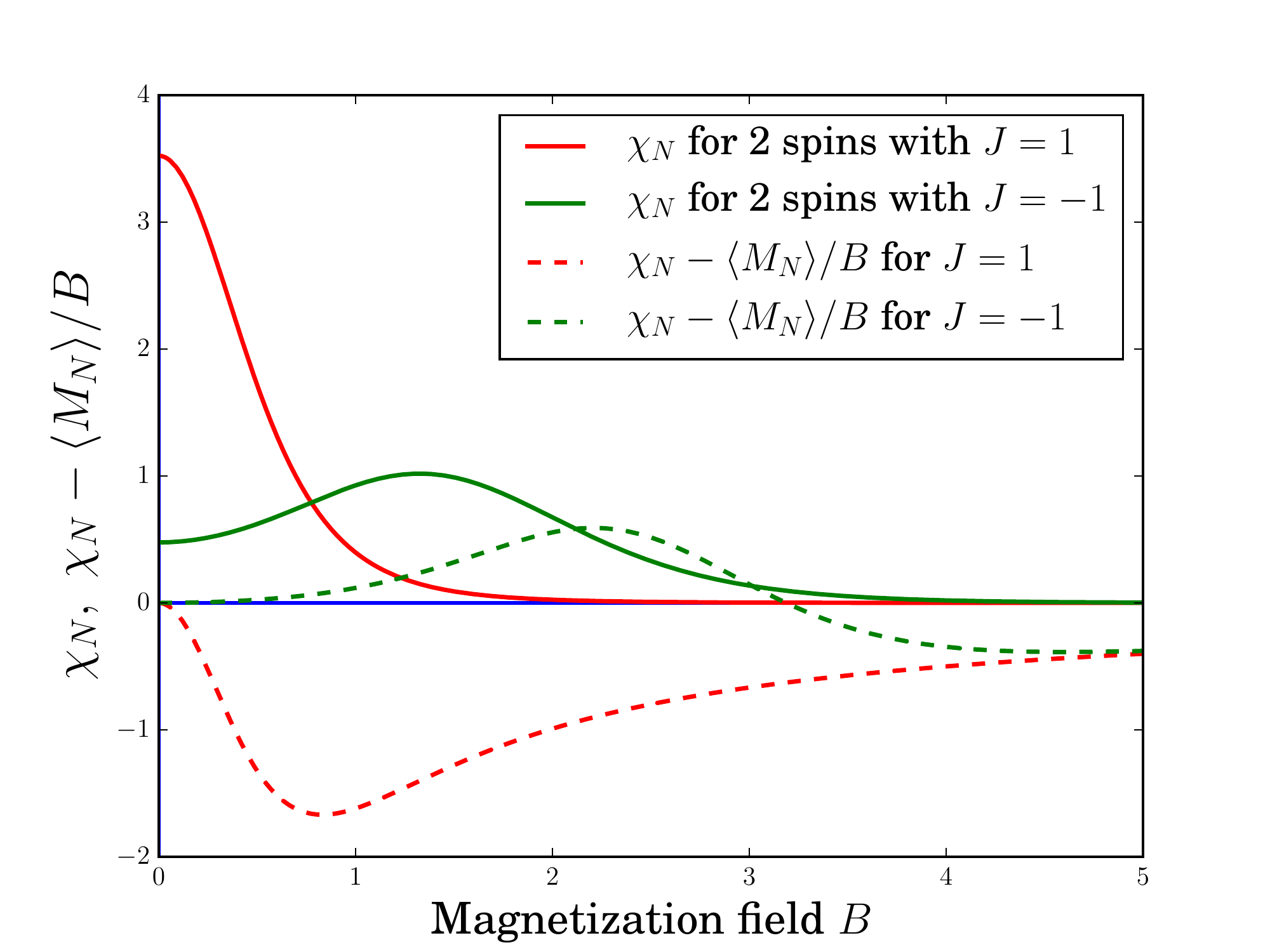}
\caption{Magnetic susceptibility $\chi_N$ (resp. $\chi_N-\langle M_N \rangle/B$) 
as a function of the magnetic field $B$ for two spins with 
ferromagnetic ($J=1$) as red solid line (resp. dashed) or 
with antiferromagnetic interaction ($J=-1$) as green solid line (resp. dashed). 
}
\label{fig:chi}
%\end{center}
\end{figure}

In order to understand this point, we plot in figure \ref{fig:chi}, the magnetic susceptibility $\chi_N$ and the quantity $\chi_N-\langle M_N \rangle/B$
for two spins with ferromagnetic or antiferromagnetic coupling. In the case of two 
spins with ferromagnetic coupling, $\chi_N$ is a monotonously decreasing function of the magnetic field $B$ on an interval 
of the form $[0,\infty[$, 
and Eq.~(\ref{chi_ineq}) holds. In contrast, for antiferromagnetic coupling, 
$\chi_N$ is non-monotonous and Eq.~(\ref{chi_ineq}) is violated in a certain range of magnetic field. 
%One can notice that $\chi_N$ is a decreasing function of $B$ on an interval 
%of the form $[0,\infty[$ for the inequality to hold, as in the ferromagnetic case.
We shall come back to this interesting observation later.  

We are now in position to generalize these results further. Let us consider an arbitrary ensemble 
%of Ising spins in a magnetic field $B$ and let $P_0(M_N)$ be the distribution of the magnetization in the absence of the field. 
of $N$ classical spins $\s=\{ \pmb{\sigma}_i \}_{i=1}^N$ taking 
discrete or continuous values such that $\pmb{\sigma}_i\in{\mathbb R}^d$ and 
$\Vert\pmb{\sigma}_i\Vert=1$ \cite{Lacoste2015a}.  The Hamiltonian of the system is assumed to be of the form
\be
H_N(\s;\B)=H_N(\s;{\bf 0})- \B \cdot \M_N(\s).
\label{Hamilt}
\ee 
By a simple calculation (see appendix A for details), one obtains
\be
\frac{\langle M_N^2 \rangle}{\langle M_N \rangle} =\frac{ \int_{M_N>0} d\M_N M_N^2  \; \cosh ( \beta \B \cdot \M_N ) \; P_{\bf 0}({\bf M_N})}{\int_{M_N>0} d\M_N M_N 
\;  \sinh ( \beta \B \cdot \M_N ) \; P_{\bf 0}({\M_N})}.
\ee
Now, we use the inequality $\tanh(x) \le x$ for $x\ge 0$, which is equivalent to 
$\sinh(x) \le x \cosh(x)$. By reporting the latter inequality into the denominator, one obtains 
\be
\frac{\langle M_N^2 \rangle}{\langle M_N \rangle} \ge \frac{ \int_{M_N>0} d\M_N M_N^2  \; \cosh ( \beta B M_N ) \; P_{\bf 0}({\M_N})}{\int_{M_N>0} 
d\M \beta B M_N^2 \;  \cosh ( \beta B M_N ) \; P_{\bf 0}({\M_N})}.
\ee
After simplifying the right hand side, we obtain Eq.~(\ref{lower-bound}) which is thus proven for any 
ensemble of classical spins with arbitrary couplings, as long as the system's Hamiltonian is given by Eq.~(\ref{Hamilt}).

This simple derivation does not work for Eq.~(\ref{inequality}), which is unfortunate because Eq.~(\ref{inequality}) is more informative than Eq.~(\ref{lower-bound}) 
- specially considering the large $B$ limit - and is a closer analog 
 to the nonequilibrium uncertainty relation \cite{Barato2015,Pietzonka2016,Gingrich2016}. For this reason, we focus 
below on Eq.~(\ref{inequality}).  
%In view of all these points, we further study Eq.~(\ref{inequality}) in the rest of this paper.

\begin{figure}[h!]
%\begin{center}
%\onefigure[scale=0.3]{figure_LD.pdf}
\onefigure[scale=0.4]{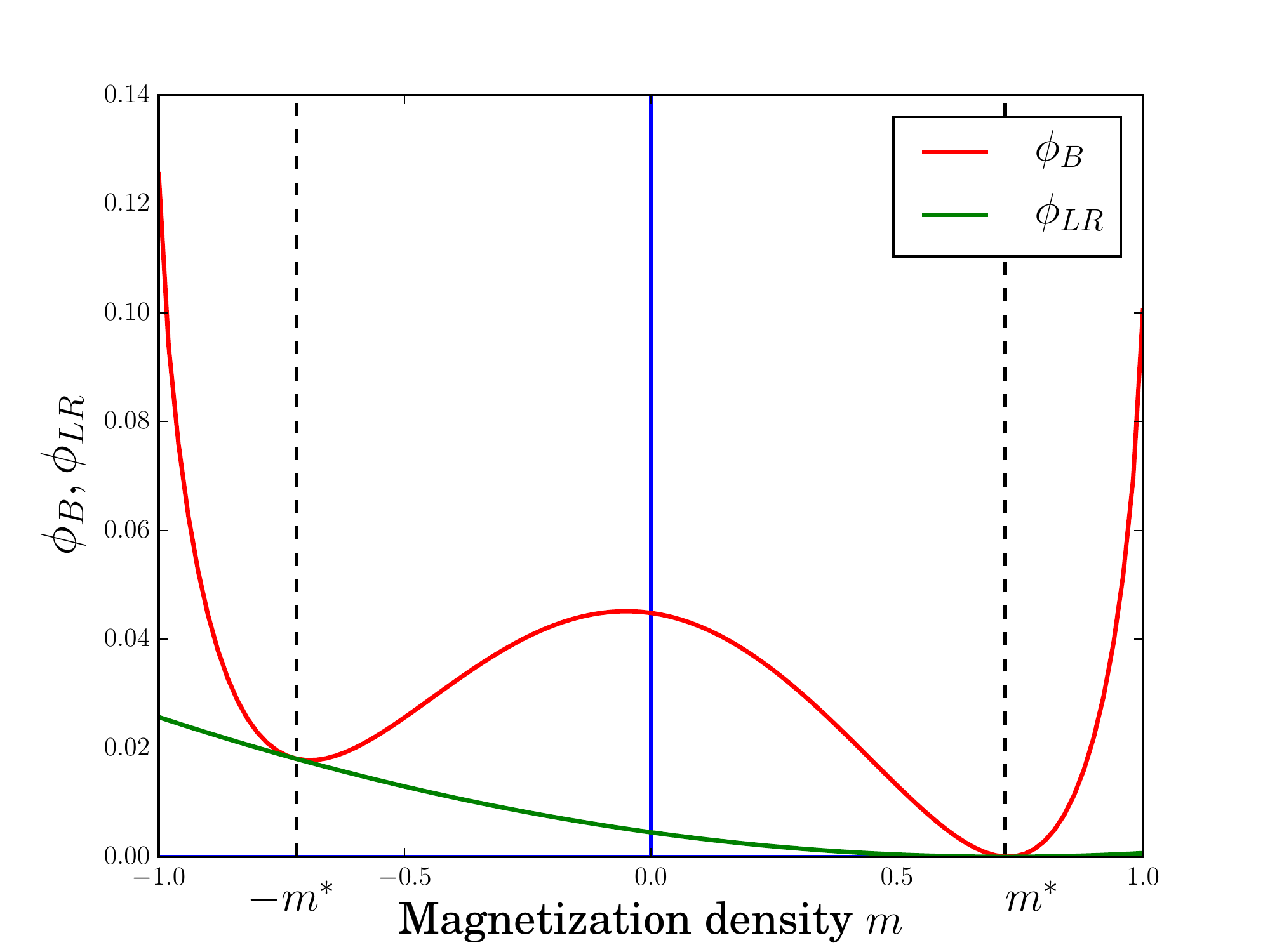}
%{\par\centering {\rotatebox{0}{\onefigure[scale=0.2]{figure0.pdf}}} 
%{\rotatebox{0}{\onefigure[scale=0.2]{figure1.pdf}}} \par} 
\caption{Large deviation function $\Phi_\B(\m)$ and its linear-response approximation $\Phi_{LR}(\m)$ 
for the {\it global} fluctuations of the magnetization density $\m$ in the 
Curie-Weiss model for a temperature $T=0.8$ (in the ferromagnetic phase) and a magnetic field $B=0.01$. The two functions 
are tangent at the points $\pm m^*$ the locations of which are shown by the two vertical dashed lines.}
\label{fig:LD}
%\end{center}
\end{figure}

Let us consider a large number of spins $N$, 
so that we can use the large deviation function \cite{Derrida2007_vol,Touchette2009}:
\be
P_{\bf B}(\M_N) \simeq e^{-N \Phi_\B(\m)}.
\label{LD-def}
\ee
Let us introduce the function
\be
\Phi_{LR}(\m)=\beta \frac{(\m - \m^*)^2 \B \cdot \m^*}{2 (m^*)^2},
\ee
where $\m^*$ is the most probable value of the magnetization which is 
such that $\Phi_\B(\m^*)=\Phi_\B'(\m^*)=0$.
This index $LR$ in $\Phi_{LR}$ indicates that this is a 
linear response regime with respect to $\B$ \cite{Gingrich2016}.
Using the fluctuation theorem of Eq.~(\ref{FT1}), it is easy to verify that $\Phi(\m)$ and $\Phi_{LR}(\m)$
 take the same value and their derivatives are equal at the two symmetrically placed points $\pm \m^*$.
To illustrate this geometrically, the two functions $\Phi_\B(\m)$ and $\Phi_{LR}(\m)$ are shown in fig. \ref{fig:LD}, 
for the particular case of the mean-field Curie-Weiss model in the ferromagnetic phase. 
In this case, note the concavity of $\Phi_\B(\m)$ in the coexistence region $-m^* \le  m \le m^*$. In contrast, this 
region becomes flat for the 2D Ising model in the limit $B \to 0$ \cite{Touchette2009}.  
%For this model, one can prove explicitly that the inequality $\Phi_\B(\m) \ge \Phi_{LR}(\m)$
%holds at all temperatures. 
%We now denote .
%Note that since $\Phi_\B(\m)$ is convex, the part of the curve between $[-m^*,m^*]$ with negative curvature 
%should be replaced by a straight line following the common tangent construction
% \cite{Touchette2009}. 

Now the variance of the order parameter is $\Phi_\B''(\m^*)=1/(N {\rm Var}(\m))$,
and $ \langle m \rangle=m^*$, because $m^*$ is unique for large enough $N$. Thus,
Eq. (\ref{inequality}) holds if    
\be
\Phi_B''(\m^*) \ge \frac{\beta B}{m^*},
\label{LD-bound2}
\ee
where the prime denotes the gradient component in the direction of $\B$. % and $m$ is the projection of $\m$ along $\B$.
Now, since $\Phi_\B(\m)=\Phi_0(\m)-\Phi_0(\m^*) - \beta \B \cdot (\m - \m^*)$ \cite{Lacoste2015a},
one has $\Phi_\B'(\m^*)=\Phi_0'(\m^*)- \beta \B=0$ and $\Phi_\B''(\m)=\Phi_0''(\m)$. Thus, 
the inequality Eq.~(\ref{LD-bound2}) is implied by the positiveness of the function
\be 
h(\m)=\Phi_0''(\m)-\Phi_0'(\m)/m,
\label{def-h}
\ee 
which can not depend on the value of the magnetic field $B$. 
%Now, $\Phi_0(\m)$ is convex from being the Legendre-Fenchel transform of another function defined in Ref. \cite{Lacoste2015a}, 
%thus $\Phi_0'(\m)$ is an increasing function of $m$. 
Further, $\Phi_0'(\m)=-\Phi_0'(-\m)$  by 
symmetry and $\Phi_0'({\bf 0})=0$. 
%Since, $\Phi_0'(\m)=\inf_\B(\Phi_\B'(\m)+\Phi_0'(\m^*))$ 
Now, when $\Phi'_0(\m)$ is convex for $m >0$ and concave for $m < 0$, $h(\m)$ is positive. 

In other words, we must have basically $\Phi'''_0(\m)>0$ for $m>0$. Now, 
it is possible to relate this condition to the one found in our earlier study of the two spins. Indeed, since $\Phi_0'(\m^*)= \beta \B$, 
by taking a derivative with respect to $B$, one obtains $\chi_N =\beta N/\Phi_0''(\m^*)$. By taking a further derivative
with respect to $B$, one finds $d\chi_N/dB=-\beta \chi_N \Phi'''_0(\m^*)/[\Phi''_0(\m^*)]^2$. Further, $\chi_N \geq 0$ by the 
fluctuation-response relation. Therefore, the 
condition $\Phi'''_0(\m)>0$ is equivalent to the condition that the susceptibility 
be a monotonously decreasing function of $B$ on the interval $B>0$, 
which is the condition found earlier in our study of the two spins case. 
To summarize, the same condition must be met for 
Eq.~(\ref{LD-bound2}) and therefore Eq.~(\ref{inequality}) to hold, both at the level of two spins or with a large number of them.

%We show below that this occurs in the Curie-Weiss model,
%but we expect this property to hold also for the 2D Ising model \cite{Touchette2009}
%and more generally for systems having a continuous phase transition when $B=0$. 

As an illustration, we can consider the Curie-Weiss model with Ising spins.   
The large deviation function of that model is \cite{Gaspard2012}:
\be
\Phi_\B(m) =  I(m) - \frac{1}{2} \beta J m^2 - \beta B m - \beta f(B).
\ee
where $f(B)$ is the Helmholtz free energy per spin and 
$I(m)$ is the classic entropy function :
\be
I(m)= \frac{1+m}{2} \ln \frac{1+m}{2} + \frac{1-m}{2} \ln \frac{1-m}{2}.
\ee
The most probable value of the magnetization, $m^*(B)$ given the magnetic field $B$, satisfies the relation $\Phi_\B'(m^*)=0$, which leads to 
the well-known self-consistent equation $m^*=\tanh(\beta J m^* + \beta B)$.
%Above the critical temperature $k_B T_c=J$, this equation has one solution, whereas below this temperature, 
%there are three solutions.
%Furthermore, since $-1 \le m \le 1$, $\Phi_B''(m) \ge \Phi_B''(0)=1-\beta J$ for all $m$.  
The function $\Phi_\B(m)$ is shown in fig. \ref{fig:LD} in the ferromagnetic phase. Note that the region $[-m^*,m^*]$ 
defines the coexistence region, where $\Phi_\B(m)$ is concave.

The function $h(m)$ introduced in Eq.~(\ref{def-h}) is    
\be
h(m) = \frac{1}{1-m^2} - \frac{\tanh^{-1}(m)}{m}.
\ee
Since it is of the form $f'(m)-f(m)/m$, 
with $f$ is convex for $m \ge 0$ and 
concave for $m \le 0$, it follows that $h(m)$ is indeed positive. 
%This can be checked explicitly since the second derivative of $f$ equals 
%$\Phi_{\bf 0}'''(m)$ given above. 
From the positiveness of $h(m)$, the inequality of Eq.~(\ref{LD-bound2}) holds, which then implies the 
bound for the fluctuations of the {\it global} order parameter given by Eq.~(\ref{inequality}).

\begin{figure}[h!]
%\begin{center}
\onefigure[scale=0.4]{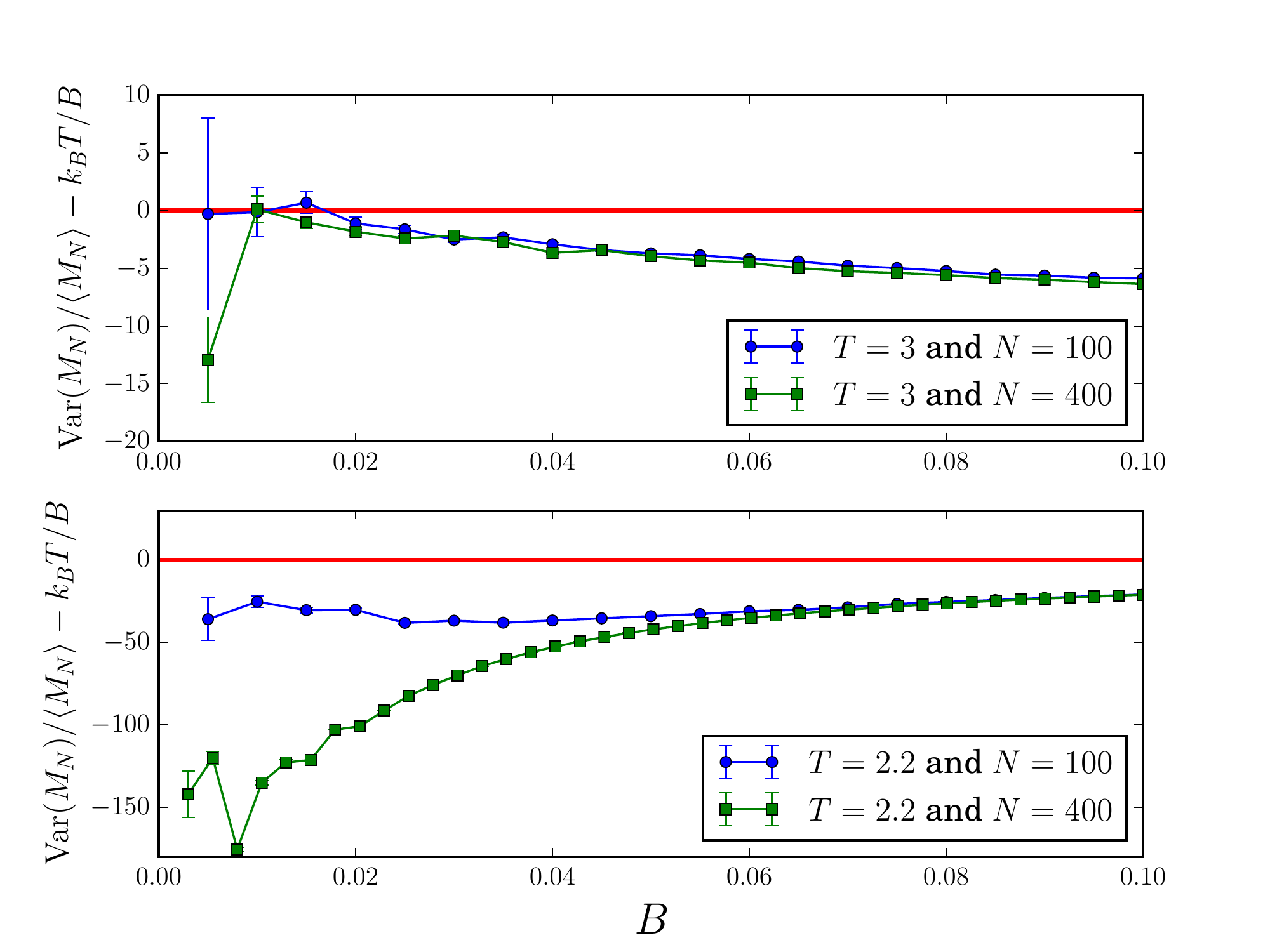}
%{\par\centering {\rotatebox{0}{\onefigure[scale=0.2]{figure0.pdf}}} 
%{\rotatebox{0}{\onefigure[scale=0.2]{figure1.pdf}}} \par} 
\caption{Difference between the left and right hand side of Eq.~(\ref{inequality}) for the 
2D-Ising model as a function of the magnetic field $B$, for two sizes $N=100$ and $N=400$. 
The temperature is $T=3.0$ (paramagnetic phase) in the upper figure and $T=2.2$ 
(ferromagnetic phase) in the lower figure.}
\label{fig:LD2}
%\end{center}
\end{figure}

For the case of the 2D Ising model, we resort to numerical simulations since we 
are not able to check directly this condition on the function $h(m)$. The results are
shown in fig.~\ref{fig:LD2}. 
In order to test this, the difference between the left hand side and the right hand side in Eq.~(\ref{inequality}) is 
plotted as a function of $B$, so that all the points should be below the red line $y=0$ according to the inequality. 
The errorbars have been estimated using the method of Ref.~\cite{Flyvbjerg1989}. These errorbars  
increase rapidly as $B \to 0$ in a system size dependent manner due to the singularity in the derivative of the free energy $F(\B)$ 
at $B=0$. We confirm that the bound holds for the paramagnetic phase (upper figure) as well as for the ferromagnetic phase (lower figure). 
It is more tight in the former case than in the latter, since the fluctuations are Gaussian in the former case.
%In contrast, we found that the inequality is not obeyed for the whole range of values of $B$ for the antiferromagnetic model.
%Most point have a positive value of the ratio variance over mean  
%magnetization appear. 
%Nevertheless, reversals of the magnetization 
%are possible corresponding to negative values of that ratio. Such reversals become very unlikely as $N$, but in 
%any case, they are allowed and they do not compromise the inequality of Eq. (\ref{inequality}). 
This analytical and numerical study of the thermodynamic bounds of Eqs.~(\ref{inequality})-(\ref{lower-bound})
represents our first main result.

We now investigate how such bounds are modified when we do not have access to 
the {\it global} order parameter, but rather to a coarse-grained or {\it local} one. 
Since the bounds are related to the fluctuation theorem, we need to generalize 
Eq.~(\ref{FT1}) for such a case. A similar situation arises out of equilibrium
due to coarse-graining \cite{Garcia-Garcia2016,Alemany2015,Michel2013,Esposito2012_vol85,Puglisi2010,Rahav2007}.
Understanding how to extract relevant information in such cases is rather pertinent experimentally even at equilibrium 
since {\it local} measurements are often the only choice, in the frequent case that the system is just too 
big to be analyzed globally.  
\begin{figure}[h!]
%\begin{center}
%\onefigure[scale=0.4,trim = 0mm 50mm 20mm 20mm, clip]{Schema}
\onefigure[width=\linewidth]{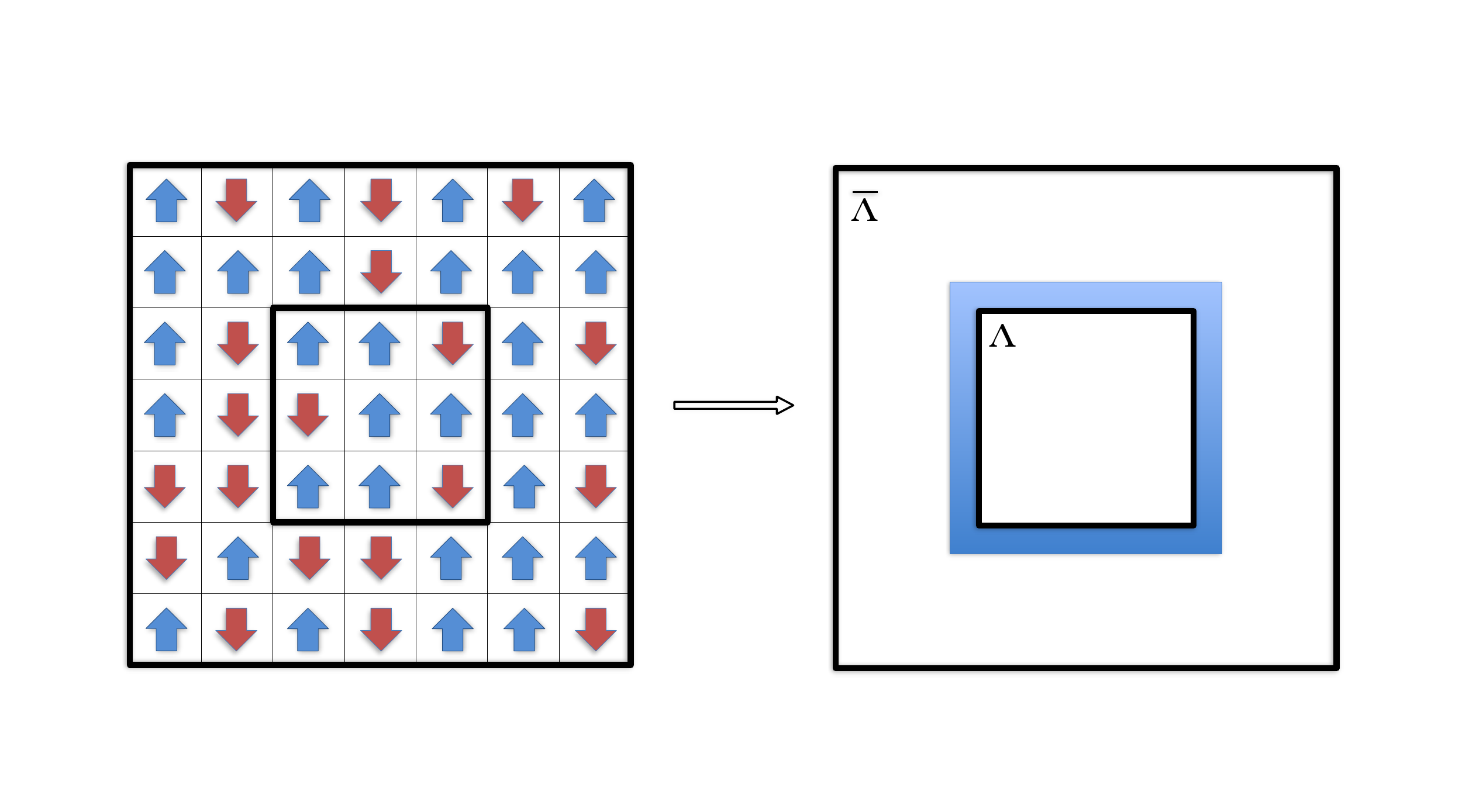}
\caption{Sketch of the magnetic system with a central square region $\Lambda$ representing the observation window. 
Spins in the complementary region $\bar{\Lambda}$ belonging to the blue area are strongly coupled to that of $\Lambda$.}
\label{fig_sketch}
%\end{center}
\end{figure}

In order to study a {\it local} version of Eq.~(\ref{FT1}), we
consider a subset of the $N$ spins containing $n<N$ spins only, $\Lambda=\{ \pmb{\sigma}_i \}_{i=1}^n$, with magnetization 
$\M_n(\s)=\sum_{i=1}^n \s_i$ as shown in fig.~\ref{fig_sketch}. 
The remaining spins $\bar{\Lambda}= \{ \pmb{\sigma}_i \}_{i=n}^N$ play the role of an ``environment'' for the spins of 
$\Lambda$.
This environment has a magnetization $\Mb_n(\s)=\sum_{i=n}^N \s_i$, so that $\M_N=\M_n + \Mb_n$.
The local equivalent of Eq.~(\ref{FT1}) is:
\be
P_\B(\M_n) = P_\B(-\M_n) \ {\rm e}^{\beta \left( 2 \B \cdot \M_n + \Gamma_\B(\M_n) \right)}.
\label{local FT1}
\ee
We have introduced the function
\begin{eqnarray}
\label{Gamma}
\Gamma_\B(\M_n) &=& k_B T \ln \langle e^{-2 \beta \B \cdot \Mb_n} | -\M_n \rangle_\B, \\
		&=& k_B T \ln \int e^{-2 \beta \B \cdot \Mb_n} {\rm Proba} (\Mb_n | -\M_n ) d^3 \Mb_n, \nonumber
\end{eqnarray}
 where ${\rm Proba} (\Mb_n | -\M_n )$ denotes the conditional probability of $\Mb_n$ given 
a magnetization $-\M_n$ for the sub-part. Thus, $\Gamma_\B(\M_n)$
is a correction factor which quantifies the failure of Eq.~(\ref{FT1}) due to the reduction of
available information in the fluctuations.
By construction, this factor must be an odd function of $\M_n$, 
 {\it i. e.}: $\Gamma_{\B}(-\M_n)=-\Gamma_{\B}(\M_n)$.

\begin{figure}[h!]
%\begin{center}
\onefigure[scale=0.4]{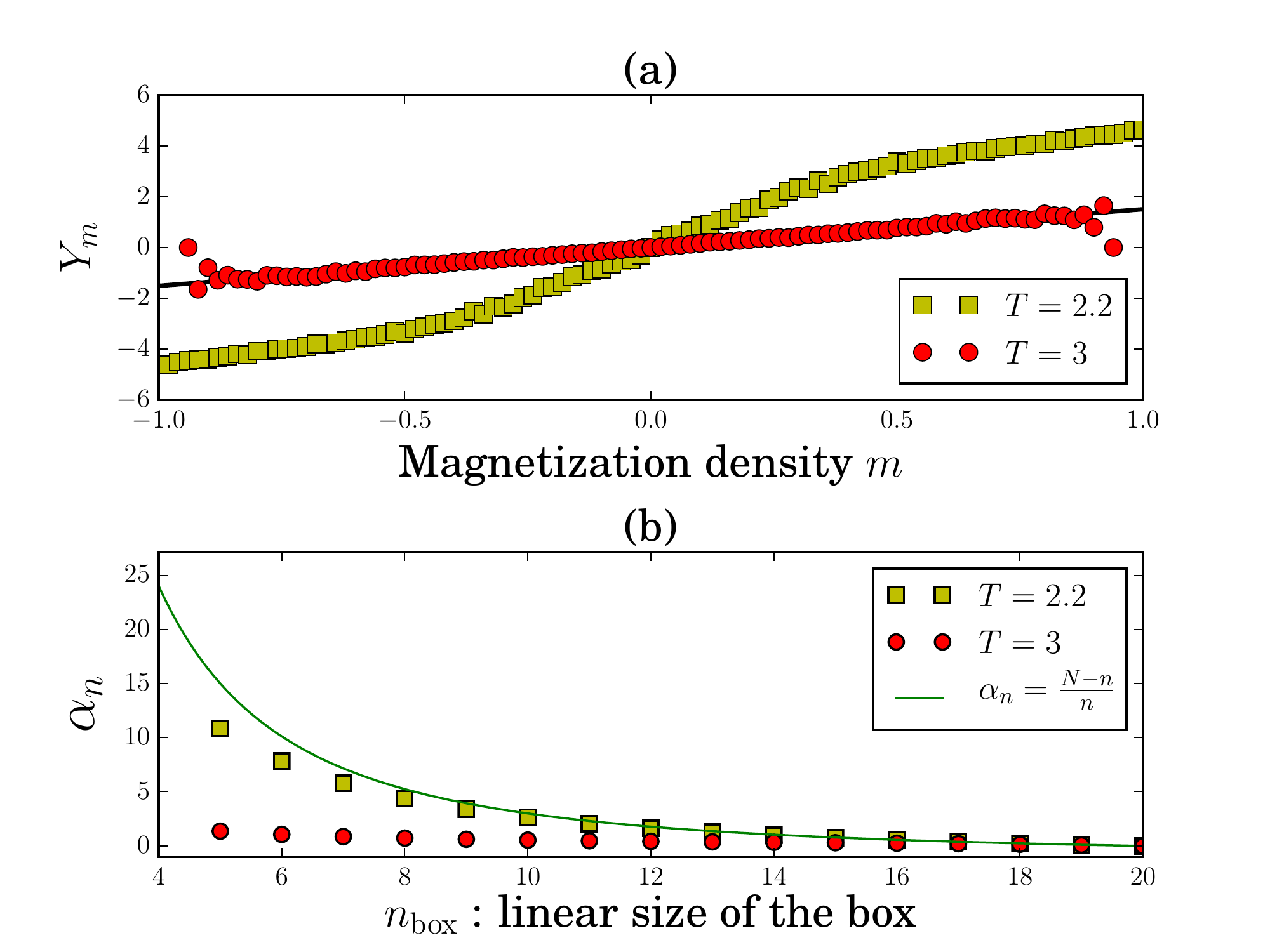}
\caption{(a) Asymmetry function $Y_m$ versus magnetization density $m$ 
for the 2D Ising model in a magnetic field, at a temperature $T=3$ (above $T_c$) or $T=2.2$ (below $T_c$). 
%While this function is always a straight line of slope one according to Eq.~(\ref{FT1}) when evaluated for the full system (dashed line), 
%it differs from it when evaluated for a sub-part (symbols). 
The order parameter is evaluated in a sub-part of $n=100$ spins among a total of $N=400$ spins, 
and the magnetic field is $B=0.01$. The critical temperature is $T_c \simeq 2.38$ for this system size.
(b) Dependence of $\alpha_n$ versus the size of the sub-part (symbols) for the same two temperatures. The solid line shows the 
dependence which is expected as the critical point is approached.}
\label{fig1}
%\end{center}
\end{figure}
In the case that all the spins of the sub-part and the rest are independent, there are no correlations between $\Lambda$ and 
 $\bar{\Lambda}$, which means ${\rm Proba} (\Mb_n | -\M_n )={\rm Proba} (\Mb_n)$. 
Using a Jarzynski like relation immediately deduced from 
Eq.~(\ref{FT1}) for the complementary part, one has $\Gamma_\B(\M_n)=\Omega_\B(\M_n)=0$. 
Therefore, the breaking of the Fluctuation relations Eq.~(\ref{FT1}) arises  
entirely from the correlations between the domains $\Lambda$ and $\bar{\Lambda}$.

In order simplify this problem, we further split the ``environment'', namely $\bar{\Lambda}$, into a subset 
of strongly correlated spins (the blue area in figure \ref{fig_sketch}), and the rest of the spins, 
which are less correlated \cite{Michel2013}.
This can be written as:
\be
\label{Langevin type}
\Mb_n=\alpha_n \M_n + \xi_n,
\ee
where we require $\xi_n$ to be uncorrelated with $\M_n$. 
In particular, this form should hold above $T_c$, where the correlation length is small and all the spins of $\bar{\Lambda}$
are uncorrelated with that of $\Lambda$ except for those at the interface between both domains.

Since $\xi_n$ is uncorrelated with $\M_n$,    
$\alpha_n$ equals the normalized co-variance between $\M_n$ and $\Mb_n$:
\be
\alpha_n=\frac{\langle \M_n \cdot \Mb_n \rangle - \langle \M_n \rangle \cdot \langle \Mb_n \rangle}{\langle \M_n^2 \rangle - \langle \M_n \rangle^2}.
\label{prop-alpha}
\ee
Then, using Eq.~(\ref{Gamma}) and Eq.~(\ref{Langevin type}), one finds a linear correction $\Gamma_\B(\M_n)=2 \alpha_n \M_n \cdot \B$.
The asymmetry function $Y_m$ defined by
\be
Y_m=\frac{1}{2 \beta B n} \ln \frac{P_\B(m)}{P_\B(-m)},  
\ee
is a straight line of slope one for the {\it global} order parameter due to Eq.~(\ref{FT1}) but becomes a straight line of slope $1+\alpha_n$ 
for the {\it local} order parameter. 
When $\alpha_n$ does not depend on the magnetization, the change of slope can be described by the 
inverse effective temperature $\beta_{eff}=\beta (1+\alpha_n)$ or by an effective magnetic field, 
similarly to the nonequilibrium case \cite{Garcia-Garcia2016}.
Since the magnetization of $\bar{\Lambda}$ acts like a field for $\Lambda$ enhancing its magnetization, $\alpha_n \ge 0$ and 
this effective temperature is smaller than $T$.
A straight asymmetry function with a slope larger than one is indeed found 
in figure \ref{fig1}a, when analyzing the fluctuations in a box of $n=100$ spins
among a total of $N=400$ spins at the temperature $T=3$.

As $T \rightarrow T_c^+$, the correlation length increases until it becomes of the order of the size of the full system.
Then, the contribution of $\xi_n$ in Eq.~(\ref{Langevin type}) should vanish 
on average, and the average magnetization density 
is $\langle \m \rangle=
\langle \M_n \rangle/n=\langle \Mb_n \rangle/(N-n)$, which implies $\alpha_n \simeq (N-n)/n$. 
Away from the critical point, $\alpha_n$ also scales as $1/n$ but the prefactor does not have such a simple form.
We have checked numerically that indeed $\alpha_n \simeq (N-n)/n$ near the critical point as shown in fig.~\ref{fig1}b. 
Using such a determination of $\alpha_n$, one could infer 
 the relative size of the observation window to the size of the large system.
In contrast, below the critical point, the 
asymmetry function of the {\it local} order parameter, has 
a sigmoidal shape as shown in fig. \ref{fig1}
when the temperature is $T=2.2$.
A similar shape is found in the case of the mean-field Curie Weiss model 
which is completely solvable analytically (see appendix B
for details of the derivation of the correction factor $\Gamma_\B(\M_n)$ for this case).

%--------MF bound ----------------------------

Let us now finally go back to our initial topic of thermodynamic bounds of the type of Eq.~(\ref{inequality}) but   
%We now turn to consequences of the modified fluctuation theorems, namely Eq.~\ref{local FT1} 
now for {\it local} fluctuations of the order parameter. 
The relevant large deviation function is defined as
\be
P_{\bf B}(\M_n) \simeq e^{-n \phi_\B(\m)},
\label{def LD-local}
\ee
for $n$ sufficiently large. Below, we use the same notation for the magnetization density $\m=\M_n/n$.
In view of the modified fluctuation theorem of Eq.~(\ref{local FT1}), the approximation  
\be
\phi_{LR}(\m)= \beta B  Y_{m^*} \frac{(\m - \m^*)^2  }{2 (m^*)^2},
\label{Phi-LR-local}
\ee
is correct by construction close to $\m=\m^*$ and has the expected value at $\m=-\m^*$ but unlike 
$\Phi_{LR}$ may not have the correct tangent at this point. 

To see precisely when this property holds, we start 
with an equivalent form of Eq.~(\ref{local FT1}), namely:
\be
\phi_\B(\m) - \phi_\B(-\m) = -2 \beta \B \cdot \m - \frac{\beta \Gamma(n \m)}{n},  
\label{FT-phi}
\ee
with $\Gamma(n \m)$ related to the asymmetry function $Y_m$ by
\be
Y_m = m + \frac{\Gamma(n \m)}{2 B n}.
\label{Ym-Gamma}
\ee
Using Eq. (\ref{FT-phi}) and the property $\phi_\B'(\m^*)=0$, one deduces that
\be
\phi_B'(-\m^*)=-2 \beta B - \beta \Gamma'(n \m^*),
\ee
while from Eq.~(\ref{Phi-LR-local}), one obtains
\be
\phi_{LR}'(-\m^*)=-2 \beta B - \beta \frac{\Gamma(n \m^*)}{n m^*}.
\ee
Therefore, we see that $\phi_B'(-\m^*)=\phi_{LR}'(-\m^*)$ if and only if  
\be
\Gamma'(n\m^*)= \frac{\Gamma(n \m^*)}{n m^*}.
\ee
From this and given that $\Gamma(0)=0$, this condition is satisfied whenever 
(i) $m^* \to 0$, which is for instance the case when $\B \to {\bf 0}$ and $T>T_c$,
(ii) the size of the sub-part goes to zero $n \to 0$, or more generally
(iii) the asymmetry function $Y_m$ is a linear function of $m$ of the form $Y_m=(1 + \alpha_n) m$,
where $\alpha_n$ is the coefficient introduced earlier.
When one of these conditions hold, the function $\phi_{LR}$ approximates $\phi_\B$ 
for all values of the magnetization, because both functions are tangent at $\m=\pm m^*$. 

In such a case, the previous derivation of the thermodynamic bound 
applies directly in terms of the effective field $B_{eff}=(1+ \alpha_n) B$, so that 
the generalization of Eq.~(\ref{inequality}) is:
\be
\frac{{\rm Var}(M_n)}{\langle M_n \rangle} \le \frac{k_B T}{B_{eff}},
\label{inequality2}
\ee
or ${\rm Var}(m)/\langle m \rangle \le k_B T/n B_{eff}$ for the magnetization density.

As a particular case, the result holds for the Curie-Weiss model. Indeed, after a straightforward calculation, the 
large deviation for the fluctuations of the {\it local} order parameter defined in 
Eq. (\ref{def LD-local}) reads \cite{Lacoste2015a},  
\ba
\phi_\B(m)&=& I(m) - \beta B m - \frac{\beta J}{2} m ^2 \frac{n}{N} 
 - \frac{N \beta f(B)}{n} \nonumber \\
 &+& \beta f(B+ J n m /N) \left( \frac{N}{n}-1 \right).
\label{explicit local LD}
\ea
This expression allows to compare the large deviation function $\phi_\B(m)$ and its linear-response approximation $\phi_{LR}(m)$
for various size ratios of the sub-part to the full system. For general values of the size ratio and  
within the ferromagnetic phase, we have checked that 
$\phi_{LR}$ is indeed not 
tangent at the point $m=-m^*$ although both functions $\phi_{LR}(m)$ and $\phi_\B(m)$ take the same value there.
When considering smaller size ratios of the sub-part to the full system or within the paramagnetic phase,  the two curves become tangent at $m=-m^*$.
In such conditions, the bound on the fluctuations of the {\it local} fluctuations, Eq.~(\ref{inequality2}) holds.

\begin{figure}[h!]
%\begin{center}
\onefigure[scale=0.4]{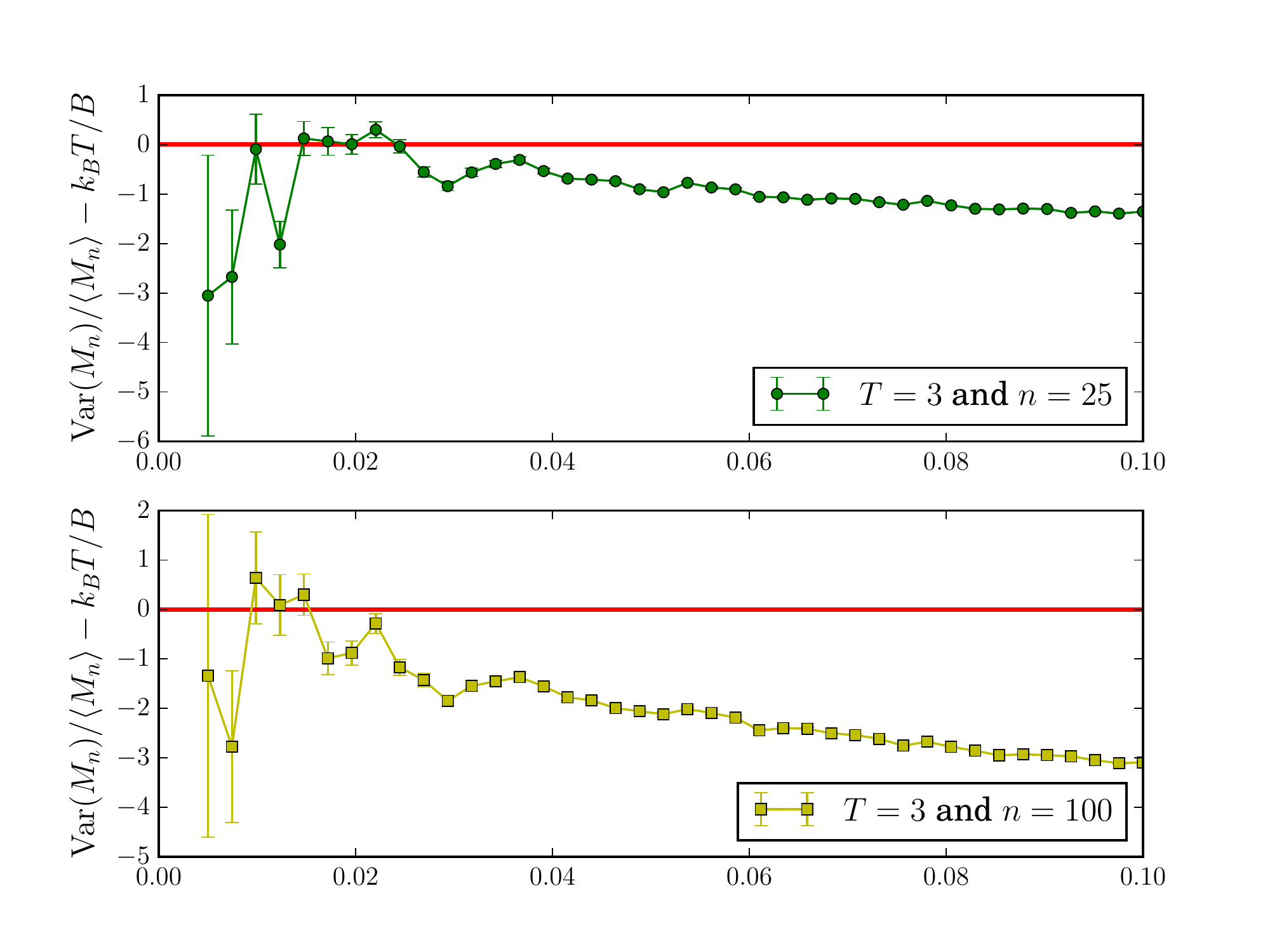}
%{\par\centering {\rotatebox{0}{\includegraphics[scale=0.2]{figure0.pdf}}} 
%{\rotatebox{0}{\includegraphics[scale=0.2]{figure1.pdf}}} \par} 
\caption{Difference between the left and right hand side of 
Eq. (\ref{inequality2}) as a function of $B$ confirming the bound for {\it local} fluctuations for the 
case of the 2D-Ising model at a temperature $T=3$ (in the paramagnetic phase) 
for two sizes of the sub-part $n=100$ and $n=25$. The values of effective magnetic field  
have been evaluated using the $\alpha_n$ of fig. \ref{fig1}.}
\label{fig:LD3}
%\end{center}
\end{figure}

For the case of the 2D Ising model, we use again numerical simulations. The results are shown in fig.~\ref{fig:LD3}, 
where the difference between the left and right hand side of 
Eq. (\ref{inequality2}) is shown, in the paramagnetic phase with the appropriate expression of $B_{eff}$ 
for two choices for the size of the sub-part $n=25$ or $n=100$. 
This verification confirms the bound for {\it local} fluctuations for this model, which represents
our second main result.
%case while it does not in the ferromagnetic phase. Nevertheless, even in that case, the data points
%admit an upper bound of the form of Eq.~(\ref{inequality2}).

To conclude, we have derived thermodynamic bounds on equilibrium 
fluctuations of {\it global} and {\it local} order parameters.
The bound for the fluctuations of a {\it global} 
order parameter is analogous to the one derived recently out of equilibrium \cite{Barato2015,Pietzonka2016,Gingrich2016}. 
In this formal analogy, the average entropy production must replaced by the magnetic field. 
This is expected since out of equilibrium, the entropy production quantifies the degree of breaking of time-reversal symmetry,
while in equilibrium, the magnetic field is responsible for the breaking of the spatial 
discrete symmetry.
%except for 
%one difference: it represents an upper bound instead of a lower bound: this is due to our parabolic approximation 
%of the large deviation 
%which is a lower bound as opposed to an upper bound in the non-equilibrium case. 

The two thermodynamic bounds contain the following trade-offs: 
Out of equilibrium, the bound imposes that reducing current fluctuations 
costs a minimal dissipation \cite{Barato2015,Pietzonka2016,Gingrich2016}; in 
equilibrium, Eq.~(\ref{inequality}) imposes that reducing order parameter fluctuations can be
achieved by increasing the magnetic field (which therefore costs some energy). 

In these two relations, fluctuations are measured by their {\it variance}. 
If we choose instead to measure fluctuations by the average of the square of the order parameter, 
the picture which emerges from Eq.~(\ref{lower-bound}) is rather different: 
such a relation can describe situations where fluctuations diverge and order can be destroyed provided 
the average magnetization scales appropriately with $B$. 
For instance, near a critical point $\langle M_N \rangle_{T_c} \sim B^{1/\delta}$  \cite{Chaikin1995_vol}, 
then Eq.~(\ref{lower-bound}) implies that $\langle M_N^2 \rangle_{T_c} \ge k_B T B^{1/\delta -1}$, which means that 
$\langle M_N^2 \rangle_{T_c}$ diverges at $B = 0$ if $\delta >1$.
%This inequality is of the Mermin-Wagner \cite{Chaikin1995_vol} type, 
%because it implies that as the magnetic field is reduced, fluctuations should increase. 
%Under some conditions which are specified by the famous theorem 
%(systems must have short range interactions, be of infinite extend and the dimension must be lower or equal than 2),
%this increase in the order parameter fluctuations can destroy  
%the long range order.

%We have shown that the derivation of the bound requires the isometric fluctuation theorems, 
%which take various forms depending on the type of symmetry breaking and order parameter 
%as reviewed in Refs.~\cite{Lacoste2014,Lacoste2015a}. 

Both in and out of equilibrium, the bounds do not follow mainly from the fluctuation theorem, 
since additional properties are needed.
In the present equilibrium case, we have seen an illustration of this with the example of the two spins. 
There, we found that Eq.~(\ref{inequality}) holds whenever the susceptibility is a monotonously decreasing function of the 
magnetic field. Considering instead a large ensemble of spins, we recovered the same condition, in the form  
of the positivity of third derivative of the large deviation function of the magnetization in the zero-field model.
Our numerical study of the 2D Ising model confirms that the inequality holds for this model at any temperature 
and system size but requires ferromagnetic interactions.

Using the formalism of Fluctuation Theorems in the presence of hidden degrees of freedom developed for 
the non-equilibrium case \cite{Garcia-Garcia2016}, we have extended the uncertainty bounds to {\it local} order parameters.
Such bounds are important because they can be tested experimentally 
more easily than their large deviation counterparts (whether at equilibrium or out of equilibrium). 
They could be used to infer the value of the symmetry breaking field,  
the relative size of the observation window with respect to the full system and possibly the nature of  
the interactions (ferromagnetic vs antiferromagnetic) using only fluctuations of the order parameter.

%%%%%%%%%%%%%%%%%%%%%%%%%%%%%%%%%%%%%%%%%%%%%%%%%%%%%%%%%%%%%%%%%%%%%%
\acknowledgments
%PG: D. Lacoste 
The authors thank P. Gaspard for a previous collaboration on which this work was built,
and U. Seifert and B. Derrida for insightful comments.

%\vskip 10pt
%%%%%%%%%%%%%%%%%%%%%%%%%%%%%%%%%%%%%%%%%%%%

%\bibliographystyle{eplbib.bst}
%\bibliography{biblio}

\newpage

%\section{Appendix A: A general inequality in the canonical ensemble} 
\begin{center}
\textbf{Appendix A: A general inequality in the canonical ensemble} 
\end{center}

%\appendix
\vskip 0.5cm
%\appendix
%\begin{widetext}
Here, we provide details for the derivations of Eqs. (3), (4) and (20) of the main text.
Let us consider a system composed of $N$ classical spins $\s=\{ \pmb{\sigma}_i \}_{i=1}^N$ taking 
discrete or continuous values such that $\pmb{\sigma}_i\in{\mathbb R}^d$ and 
$\Vert\pmb{\sigma}_i\Vert=1$.  The Hamiltonian of the system is assumed to be of the form
\be
H_N(\s;\B)=H_N(\s;{\bf 0})- \B \cdot \M_N(\s)
\label{Hamilt}
\ee 
where $\B$ is the external magnetic field and the order parameter is the magnetization 
\be
\M_N(\s)=\sum_{i=1}^N \s_i \, .  
\label{Magnet}
\ee
We suppose that the system is at equilibrium in the Gibbsian canonical distribution at the inverse
 temperature~$\beta$, and we introduce 
\be
\mu_{\bf B}(\s)= \frac{1}{Z_N(\B)} \, {\rm e}^{- \beta H_N(\s;\B)} ,
\label{canonical}
\ee
where $Z_N(\B)=\sum_{\pmb{\sigma}} {\rm e}^{- \beta H_N(\s;\B)}$ is the classical partition function such that the distribution is normalized to unity: $\sum_{\pmb{\sigma}}\mu_{\bf B}(\s)=1$.

Let us also define the probability density $P_\B(\M)$ that the magnetization takes the value $\M=\M_N(\s)$ as
\be
P_\B(\M) \equiv \langle \delta\left[\M-\M_N(\s)\right]\rangle_\B
\ee
where $\delta(\cdot)$ denotes the Dirac delta distribution and $\langle \cdot\rangle_\B$ the statistical average over Gibbs' canonical measure (\ref{canonical}). 
The same distribution in the absence of the field, is denoted $P_{\bf 0}({\bf M})$, and is related to $P_\B(\M)$ by
\be
P_\B(\M) = \frac{Z_N({\bf 0})}{Z_N({\bf B})}\; {\rm e}^{\beta \B \cdot \M} \; P_{\bf 0}({\bf M}).
\label{GI}
\ee
This probability density $P_\B(\M)$  is a function of the vectorial magnetization $\M\in{\mathbb R}^d$ and it 
is normalized according to
\be
\int d\M \, P_\B(\M) = 1 \, .
\label{normalization}
\ee
We also define $M_z$ to be the component of $\M$ along the direction of $\B$, and $B$ is the amplitude of $\B$, so that $\B \cdot \M=B M_z$.

Then, using Eq. (\ref{GI})
\ba 
\langle \M^2 \rangle &=& \int d\M \M^2 P_\B(\M)  \nonumber \\
&=& \int_{M_z>0} d\M \M^2  \; 2 \cosh ( \beta \B \cdot \M ) \; P_{\bf 0}({\bf M}) \frac{Z_N({\bf 0})}{Z_N({\bf B})}. \nonumber
\ea 
Similarly, 
\ba
\langle M_z \rangle &=& \int d\M M_z P_\B(\M) \nonumber \\
&=& \int_{M_z>0} d\M M_z \; 2 \sinh ( \beta \B \cdot \M ) \; P_{\bf 0}({\bf M}) \frac{Z_N({\bf 0})}{Z_N({\bf B})}. \nonumber
\label{Mz}
\ea
We are interested in the Fano factor like ratio
\be
\frac{\langle \M^2 \rangle}{\langle M_z \rangle}=\frac{ \int_{M_z>0} d\M \M^2  \; \cosh ( \beta \B \cdot \M ) \; P_{\bf 0}({\bf M})}{\int_{M_z>0} d\M M_z \;  \sinh ( \beta \B \cdot \M ) \; P_{\bf 0}({\bf M})}.
\ee
Now using $\M^2 \ge M_z^2$, we obtain
\be
\frac{\langle \M^2 \rangle}{\langle M_z \rangle} \ge \frac{\langle M_z^2 \rangle}{\langle M_z \rangle}  = \frac{ \int_{M_z>0} d\M M_z^2  \; \cosh ( \beta \B \cdot \M ) \; P_{\bf 0}({\bf M})}{\int_{M_z>0} d\M M_z \;  \sinh ( \beta \B \cdot \M ) \; P_{\bf 0}({\bf M})}.
\ee

Then, using $\tanh(x) \le x$ for $x\ge 0$, which is equivalent to $\sinh(x) \le \cosh(x) x$ in the denominator, we obtain
\be
\frac{\langle M_z^2 \rangle}{\langle M_z \rangle} \ge \frac{ \int_{M_z>0} d\M M_z^2  \; \cosh ( \beta B M_z ) \; P_{\bf 0}({\bf M})}{\int_{M_z>0} d\M \beta B M_z^2 \;  \cosh ( \beta B M_z ) \; P_{\bf 0}({\bf M})}.
\ee
Therefore, we obtain our result, namely Eq. (3) of the main text (where $M_z$ is simply denoted $M_N$):
\be
\frac{\langle \M^2 \rangle}{\langle M_z \rangle} \ge \frac{\langle M_z^2 \rangle}{\langle M_z \rangle}  \ge \frac{1}{\beta B}.
\ee
Note that this result holds generally for any magnetic system described by Eq. (\ref{Hamilt}): there are no conditions on the number of spins or on the 
couplings between the spins.

The fluctuation-response relation given above Eq. (4) of the main text is obtained by taking the derivative of Eq. (\ref{Mz}) with respect to the magnetic field $B$.
This classic calculation leads to:
\be
\frac{d \langle M_z \rangle}{dB}= \beta \left( \langle M_z^2 \rangle - \langle M_z \rangle^2 \right)= \beta {\rm Var} M_z.
\ee
This quantity is introduced in the main text above Eq. (4) as the magnetic susceptibility $\chi_N$ and 
together with Eq. (3), it leads to Eq. (4) of the main text.

\vskip 1cm
\begin{center}
\textbf{Appendix B: Asymmetry function for the fluctuations of the local order parameter in the mean-field Curie Weiss model} 
\end{center}

%\appendix
\vskip 0.5cm

%---------------------MF------------------

Here, we provide derivation of the correction factor defined in Eq.~(20) of the main text for the  
mean field Curie-Weiss model with Ising spins.
The distribution of the {\it local} order parameter is
\be
P_{\bf B}(\M_n)=\frac{1}{Z_N({\bf B})} \sum_{\s} {\rm e}^{\frac{\beta J}{2N}\, \M_N^2 +\beta \B \cdot \M_N}  \\
 \delta\left[\M_n-\M_n(\s)\right], 
\label{P(M)}
\ee
where the sum is taken over configurations of the full system. After separating the contribution of spins 
belonging to $\Lambda$ and to $\bar{\Lambda}$, one obtains 
\be
P_{\bf B}(\M_n)=\frac{Z_{\bar{\Lambda}} (\B + J  \M_n/N)} {Z_N({\bf B})}  {\rm e}^{\frac{\beta J}{2N}\, \M_n^2 +\beta \B \cdot \M_n} C_n(\M_n),
\ee
where $Z_{\bar{\Lambda}}(\B + J \M_n/N)$ is the partition function of an ensemble of $N-n$ spins of $\bar{\Lambda}$, 
subject to the effective magnetic field $\B_{eff}=\B - J \M_n/N$ and interacting with the 
effective coupling constant $J_{eff}=J (N-n)/N$. The function $C_n(\M_n)$ is the number 
of microstates with a given value of local magnetization $\M_n$.
Since this function only depends on $|\M_n|$, one obtains the correction factor defined in Eq.~(20) of the main text:
\be
\label{Gamma_MF}
\Gamma_\B(\M_n)=F_{\bar{\Lambda}}(\B - J \M_n/N) - F_{\bar{\Lambda}}(\B + J \M_n/N),
\ee
where $F_{\bar{\Lambda}}$ is the mean-field free energy associated with  
the partition function $Z_{\bar{\Lambda}}$.

%%%%%%%%%%%%%%%%%%%%%%%%%%%%%%%%%%%%%%%%%%%%%%%%%%%%%%%%%%%%%%%%%%%%%%
%\section*{Acknowledgments}
%The authors thank D. Gaspard for helpful advices in power expansions with Mathematica \cite{W88}.

%\end{widetext}
\end{document}